\documentclass[hidelinks,12pt]{article}
\usepackage{bm}
\usepackage{amsmath, amsthm, amssymb}
\usepackage{setspace}
\usepackage{multirow}
\usepackage{graphicx}
\usepackage{natbib}
\usepackage{hyperref}
\usepackage{mathrsfs}
\usepackage{mdwlist}
\usepackage{rotating}
\usepackage{arydshln}
\usepackage{pdflscape}

\hypersetup{
    colorlinks=false,
    pdfborder={0 0 0},
}

\newcommand{\diag}{\mathop{\mathrm{diag}}}

\newcommand{\ts}{^{\sf T}}

\newcommand{\X}{{\bf X}}
\newcommand{\B}{{\bf B}}

\newcommand{\beq}{\begin{equation}}
\newcommand{\eeq}{\end{equation}}

\newcommand{\captionfonts}{\footnotesize}
\makeatletter
\long\def\@makecaption#1#2{
  \vskip\abovecaptionskip
  \sbox\@tempboxa{{\captionfonts #1: #2}}
  \ifdim \wd\@tempboxa >\hsize
    {\captionfonts #1: #2\par}
  \else
    \hbox to\hsize{\hfil\box\@tempboxa\hfil}
  \fi
  \vskip\belowcaptionskip}
\makeatother

\begin{document}

\title{Bankruptcy Prediction of Small and Medium Enterprises Using a Flexible Binary Generalized Extreme Value Model}
\author{Raffaella Calabrese$^{1}$, Giampiero Marra$^2$ and Silvia Angela Osmetti$^3$\\
\small $^1$Essex Business School, University of Essex\\ \small
Wivenhoe Park, Colchester CO4 3SQ, UK\\ \small \texttt{rcalab@essex.ac.uk} \\
\small $^2$Department of Statistical Science, University College London, \\ \small Gower Street, London WC1E 6BT, U.K.\\ \small \texttt{giampiero.marra@ucl.ac.uk} \\
\small $^3$Department of Statistical Sciences, Universit\`{a} Cattolica del Sacro Cuore di Milano \\ \small Largo Gemelli 1, 20123 Milano, Italy \\ \small  \texttt{silvia.osmetti@unicatt.it}
}

\maketitle

\begin{abstract}

We introduce a binary regression accounting-based model for bankruptcy prediction of Small and Medium Enterprises (SMEs). The main advantage of the model lies in its predictive performance in identifying defaulted SMEs. Another advantage, which is especially relevant for banks, is that the relationship between the accounting characteristics of SMEs and response is not assumed a priori (e.g., linear, quadratic or cubic) and can be determined from the data. The proposed approach uses the quantile function of the generalized extreme value distribution as link function as well as smooth functions of accounting characteristics to flexibly model covariate effects. Therefore, the usual assumptions in scoring models of symmetric link function and linear or pre-specified covariate-response relationships are relaxed. Out-of-sample and out-of-time validation on Italian data shows that our proposal outperforms the commonly used (logistic) scoring model for different default horizons.

\end{abstract}

\vspace{0.4cm}

\section{Introduction}

A significant innovation of the Basel II regulatory framework \citep{Basel:2005} is the greater use of risk assessments provided by banks' internal systems as inputs to capital calculations. Based on the internal ratings-based approach of the revised framework, banking institutions are allowed to use their own measures as input for their minimum regulatory capital calculations. The main input for these calculations is the probability of default forecasted one year ahead. Therefore, in many credit risk models, such as CreditMetrics \citep{Gupton:1997}, CreditRisk+ \citep{Credit:1997} and CreditPortfolioView \citep{Wilson:1998}, default probabilities are essential input parameters. Even if default risk could be estimated for different kinds of loans, i.e. corporate loans, those for Small and Medium Enterprises (SMEs), and the creation of a rational and comprehensive policy for them, play a central role in the European Union (EU) economy (see Small Business Act for Europe, 2008, Brussels: European Commission, http://ec.europa.eu/enterprise/entrepreneurship/docs/sba/SBA$\_$IA). Banks have realized that small and medium sized companies are a distinct kind of clients with needs and peculiarities which require specific risk management tools and methodologies \citep[e.g.,][]{Altman:2006, Dietsch:2004, Saurina:2004}.

SMEs are important for the economic system of many countries. Since lending to SMEs is riskier than lending to large corporations \citep{Altman:2006}, banks should develop credit risk models that are specific to SMEs. This generated a lot of interest in the community, which resulted in a number of solutions \citep[e.g.,][]{Altman:2006,Berger:2002,Ciampi:2008,Fantazzini:2009,Lin:2012,Saurina:2004}. We propose a binary generalized extreme value additive (BGEVA) model for predicting SMEs defaults. Specifically, since employing a symmetric link function for a rare event such as default may be problematic \citep[see][for the use of the logit link]{King:2001}, we consider the quantile function of a generalized extreme value (GEV) random variable as link function \citep{Calabrese:2013,Wang2010}. Moreover, because the assumption of a linear or pre-specified (e.g., quadratic or cubic) relationship between the accounting characteristics of SMEs and default is rarely satisfied in practice, we use penalized regression splines to flexibly determine covariate effects from the data \citep[e.g.,][]{Ruppert:2003,Wood:2006}. We apply the BGEVA model, and its traditional competitors, to data on 50,160 Italian SMEs for the period $2006-2011$, an important time horizon since it includes the financial crisis of 2008. Our empirical analysis shows that the relationship between accounting characteristics and default is not linear and that our proposal outperforms the commonly used scoring model in terms of predictive performance.

The contribution of this article is twofold. From a methodological point of view, we present a penalized likelihood estimation framework for a binary regression model where the link function is allowed to be asymmetric and the functional dependence of the response on continuous covariates is flexibly determined from the data. Importantly, the methods discussed in this article are implemented in the \texttt{R} package \texttt{bgeva} \citep{bgeva}, which makes it feasible for practitioners to fit BGEVA models. We elected to follow a frequentist approach because it can particularly appeal to researchers already familiar with traditional frequentist techniques and usually has the advantage of being computationally fast.

From an applied perspective, we improve on the prediction results obtained using classic alternatives. This is crucial for financial institutions and the economic system. If banks can discriminate better between defaulting and non-defaulting SMEs, then the credit system may become more efficient. This is pivotal for small businesses since their financial structure typically depends heavily on financial institutions that provide external funding. This means that shocks to the banking system can have a significant impact on the credit supply to small businesses. As a consequence, SMEs may be subject to funding problems and credit access may be problematic during recession periods in financial markets. As this topic is pivotal in the EU and USA \citep{Berger:2006,Pederzini:2012}, we would expect our proposal to be generally useful for analyzing the SMEs of many countries.

The paper is organized as follows. The next section provides a review of the main literature on scoring models for SMEs. In Sections \ref{LogMod} and \ref{newM}, we discuss the main drawbacks of the commonly used (logistic) regression for credit default applications and introduce the BGEVA scoring model. Section \ref{EmpricalCase} shows the results obtained from applying the traditional and proposed approaches to data on Italian SMEs. The last section is devoted to the conclusions.

\section{Literature review\label{LRew}}

Credit risk models which can separate defaulting and non-defaulting firms as well as predict corporate bankruptcy can be classified into two groups: market-based models and accounting-based models. The majority of them belong to the first group and are based on structural and reduced-form approaches \citep{Merton:1974,Jarrow:1995}. Since these models make use of capital market information, which is not available for small firms owing to their information opacity and unattainability, we focus on the second group which uses accounting variables of firms from financial statements. More details on credit risk assessment can be found in \cite{Hand:1997}.

\cite{Altman:1968} was the first to introduce an accounting-based model for estimating the default probabilities of firms. This was achieved calculating his well-known Z-Score using a standard discriminant model. Since then, different methodologies, such as linear regression, logistic regression, classification trees and neural networks have been employed for credit risk assessment \citep{Thomas:2004}. The widely used model is logistic regression \citep{Altman:2006,Lin:2012}, which shows, however, some important drawbacks for default prediction. Since default can be viewed as a rare event (because the number of defaults in a sample is typically very small) and the logit link function is symmetric, the default probability could be underestimated \citep{Calabrese:2013}. Moreover, the bias of maximum likelihood parameter estimators for logistic regression is amplified in rare event studies \citep{King:2001}.

Accounting-based models can be applied to different kinds of firms. As mentioned in the introduction, because of the higher risk of lending to SMEs than to other firms, many scholars have focused on developing tools specifically designed for SMEs. In Italy, SMEs form the $99.9\%$ of the firms (see the SBA fact sheet 2012 for Italy). These employ around the $81\%$ of the individuals in the workforce and contributed for the $68.3\%$ of the Italian Added Value in 2011 \citep{European:2012}. More generally, SMEs play a pivotal role in many countries. For example, \cite{Altman:2006} analyzed 2,010 US SMEs for the period $1994-2002$. They stressed the importance of using a model designed for assessing the rating of an SME, instead of employing a generic model built for both SMEs and large corporations. According to \cite{Altman:2006}, using different models for small/medium enterprises and large firms may lead banks to lower the required capital as established by Basel II \citep{Basel:2005}. In a later work, \cite{Altman:2010} significantly improved the default prediction power of risk models for UK SMEs by using explanatory variables such as legal action by creditors to recover unpaid debts, company filing histories and comprehensive audit report/opinion. \cite{Lin:2012} analyzed UK SMEs adopting different definitions of a failing small business. Specifically, for 429 SMEs, whose performance at the end of 2004 and financial ratios from 2001 were observed, they investigated the impacts that several default definitions have on the choice of predictor variables and model's predictive accuracy.

\section{Main drawbacks of the traditional credit scoring model\label{LogMod}}

The most commonly used model for credit scoring applications is logistic regression \citep[e.g.,][]{Altman:2006,Becchetti:2002,Lin:2012,Zavgren:1998}. Let $y_i$ be a binary response which describes the event default $(y_i=1)$ and non-default $(y_i=0)$ for the $i^{th}$ SME. The logistic model can be written as
\begin{equation}
\label{logit}
logit(PD_i)=\ln \left ( \frac{PD_i}{1-PD_i}\right) =\alpha+\sum_{j=1}^p\beta_jx_{ji}=\eta_i, \ \ j=1,2,...,p, \ \ i=1,2,...,n,
\end{equation}
where $PD_i$ is the default probability, $\alpha$ is an intercept, the $x_{ji}$ are $p$ financial and economic continuous variables, the $\beta_j$ are regression coefficients and $n$ represents the sample size. Details on estimation methods and inferential procedures can be found in \cite{McCullagh:1989}.

There are two main drawbacks associated with model (\ref{logit}). First, $PD_i$ may be underestimated. Since the number of defaults in a sample is usually very small \citep[e.g.,][]{Kiefer:2010,Lin:2012}, default can be viewed as a rare event. Hence, the use of the logit link function may not be appropriate because of its symmetry around $0.5$, which implies that the response curve, $PD_i=1/\{1+\exp(-\eta_i)\}$, approaches zero at the same rate as it approaches one. This is not ideal as the characteristics of defaults are more informative than those of non-defaults and as a consequence $PD_i$ will be underestimated. This suggests using an asymmetric link function as in \cite{Calabrese:2013} and \cite{Wang2010}. 

Second, model (\ref{logit}) assumes a linear relationship between the accounting characteristics of SMEs and response \citep[e.g.,][]{Chuang:2009, Gestel:2005,Huang:2006,Lee:2005,Ong:2005}. One could easily include quadratic or cubic terms to relax the assumption of linearity but such a procedure would require choosing a priori the order of the polynomial function for each covariate and it is not in general recommended \citep[e.g.,][]{MR10,Wood:2006}. This issue can mask possibly interesting non-linear patterns which can help improve our understanding of the underlying covariate-response relationships and perhaps improve the prediction accuracy of the scoring model as well. This calls for an approach which can flexibly determine covariate effects from the data (see, for instance, \cite{Berg:2007} who employed a logistic additive model).

The next section describes a methodology that can overcome the two aforementioned model (\ref{logit})'s drawbacks by blending the strengths of using an asymmetric GEV link function \citep{Calabrese:2013,Wang2010} and an additive predictor \citep{Berg:2007}.

\section{The BGEVA model \label{newM}}

Since the percentage of defaults is typically very low, even for SMEs, the defaulters' characteristics are more informative than those of non-defaulters \citep{Berg:2007,Kiefer:2010,Lin:2012}. This means that defaulters' features are better represented by the tail of the response curve for values close to one, which can be modeled using a GEV random variable \citep{Kotz:2000,Falk:2010}. Because our focus is on defaulters, as in \cite{Calabrese:2013}, we exploit the quantile function of a GEV random variable and specify the link function
\begin{equation}
\label{linkgev}
\frac{\left[-\ln (PD_i)\right]^{-\tau}-1}{\tau},
\end{equation}
where $\tau \in \Re$ is the tail parameter. Hence, in (\ref{logit}) we replace $logit(PD_i)$ with (\ref{linkgev}). Since a GEV link can be asymmetric, underestimation of the default probability may be overcome. As discussed, for instance, in \cite{Calabrese:2013}, depending on the value of $\tau$, several special cases can be recovered; e.g., when $\tau\rightarrow 0$ the GEV random variable follows a Gumbel distribution and its cumulative distribution is the log-log function \citep{Agresti:2002}.

Moving on to the right hand side of equation (\ref{logit}), as explained in the previous section, the assumption of linear or pre-specified covariate-response relationships may be restrictive in applications. Therefore, we replace $\alpha+\sum_{j=1}^p\beta_jx_{ji}$ with $\alpha+\sum_{j=1}^p f_j(x_{ji})$, where the $f_j(\cdot)$ are unknown one-dimensional smooth functions of the continuous covariates $x_{ji}$. The smooth terms are represented using the regression spline approach \citep{EM96,Ruppert:2003,Wood:2006}. Specifically, $f_j(x_{ji})$ is approximated by a linear combination of known (e.g., cubic or thin plate regression) spline bases, $b_k(x_{ji})$, and unknown regression parameters, $\gamma_{jk}$. That is, $f_j(x_{ji})=\sum_{k=1}^{K_j} \gamma_{jk} b_k(x_{ji})$, where $K_j$ is the number of basis functions. In other words, calculating $b_k(x_{ji})$ for each observation point and $k$ yields $K_j$ curves encompassing different degrees of complexity which multiplied by some real valued parameters $\gamma_{jk}$ and then summed give an estimated curve for the smooth component (see \cite{Ruppert:2003} for a more detailed introduction). Smooth terms are typically subject to identifiability centering constraints such as $\sum_{i=1}^n f_j(x_{ji})=0 \ \ \forall j$ \citep{Wood:2006}.

The use of both a GEV link and smooth components leads to
\begin{equation}
\label{SEMGEVA}
\frac{\left[-\ln (PD_i)\right]^{-\tau}-1}{\tau}=\alpha+\sum_{j=1}^pf_j(x_{ji})=\eta_i,
\end{equation}
where $\eta_i$ can be written as $\textbf{B}_i\ts\bm\delta$ with $\textbf{B}_i\ts=[1,b_1(x_{1i}),\ldots,b_{K_1}(x_{1i}),\ldots,b_1(x_{pi}),\ldots,b_{K_p}(x_{pi})]$ and $\bm\delta\ts=(\alpha,\gamma_{11},\ldots,\gamma_{1K_1},\ldots,\gamma_{p1},\ldots,\gamma_{pK_p})$. This represents the BGEVA model that will be employed in Section \ref{EmpricalCase}. For our case study, a more general additive predictor was not deemed to be required. It is worth mentioning, however, that the \texttt{R} package \texttt{bgeva} \citep{bgeva} supports the inclusion, for instance, of parametric terms (i.e., binary and categorical predictors), of terms obtained by multiplying one or more smooth components by some predictor(s), and of smooth functions of two or more continuous covariates (see \cite{Wood:2006} for details on these alternative specifications). Of course, $\textbf{B}_i$ and $\bm\delta$ would have to be modified accordingly but this is a minor change. Finally, note that model (\ref{SEMGEVA}) is only defined for those observations for which $1+\tau\eta_i\geq 0$ \citep[e.g.,][]{Calabrese:2013}.

\subsection{Parameter estimation}

In model (\ref{SEMGEVA}), replacing the smooth terms with their regression spline expressions yields essentially a parametric model whose design matrix includes spline bases. This means that a BGEVA model can be estimated by maximum likelihood (ML). However, classic ML estimation of (\ref{SEMGEVA}) is likely to result in smooth function estimates that are too rough to be useful for empirical analysis \citep[e.g.,][]{Ruppert:2003}. This issue can be overcome by penalized ML, where the use of penalty matrices allows us to suppress that part of smooth term complexity which has no support from the data \citep[e.g.,][]{Ruppert:2003,Wood:2006}. Specifically, each smooth has an associated penalty, $\bm\gamma_j\ts \textbf{S}_j \bm\gamma_j$, where $\bm\gamma_j\ts=(\gamma_{j1},\ldots,\gamma_{jK_j})$ and $\textbf{S}_j$ is a positive semi-definite square matrix of known coefficients measuring the roughness of the $j^{th}$ smooth component; for instance, the second-order roughness measure for a univariate spline penalty evaluates $\int \{f''(x_j)\}^2 dx_j$. The formulas for the $b_k(x_{ji})$ and $\textbf{S}_j$ depend on the type of spline basis employed and we refer the reader to \cite{Ruppert:2003} and \cite{Wood:2006} for these details.

For a fixed value of $\tau$, the BGEVA model is estimated by maximization of
\begin{equation}
\ell_p(\bm\delta)=\ell(\bm\delta) + \frac{1}{2}\sum_{j=1}^p \lambda_j \bm\gamma_j^T\textbf{S}_j\bm\gamma_j \ \  \text{w.r.t.} \ \  \bm\delta,
\label{pen.l}
\end{equation}
where
$$
\ell(\bm\delta)=\sum_{i=1}^n-y_i(1+\tau\eta_i)^{-1/\tau}+(1-y_i)\ln\{1-\exp[-(1+\tau\eta_i)^{-1/\tau}]\}, \ \eta_i=\textbf{B}_i\ts\bm\delta.
$$
Penalized log-likelihood (\ref{pen.l}) is essentially maximized iterating
\beq
 \hat{\bm\delta}^{[a+1]}=\hat{\bm\delta}^{[a]}+ (\bm{\mathcal{J}}^{[a]}-\textbf{S}_{\bm\lambda})^{-1} (\textbf{S}_{\bm\lambda}\hat{\bm\delta}^{[a]}-\bm{\mathcal{U}}^{[a]})
\label{Fish}
\eeq
until convergence, where $a$ is the iteration index, $\textbf{S}_{\bm\lambda}=\diag(0,\lambda_{1}\textbf{S}_1,\ldots,\lambda_{p}\textbf{S}_p)$ (when the additive predictor is specified as in (\ref{SEMGEVA})),
$$
\bm{\mathcal{U}}(\bm\delta)=\frac{\partial \ell(\bm\delta)}{\partial \bm\delta}=\sum_{i=1}^n \left\{y_i(1+\tau\eta_i)^{-1-\frac{1}{\tau}}+(1-y_i)\frac{\exp[-(1+\tau\eta_i)^{-1/\tau}](1+\tau\eta_i)^{-1-\frac{1}{\tau}}}{1-\exp[-(1+\tau\eta_i)^{-1/\tau}]}\right\}\B_i
$$
and
\begin{eqnarray*}
\bm{\mathcal{J}}(\bm\delta)&=&\frac{\partial^2 \ell(\bm\delta)}{\partial \bm\delta \partial \bm\delta\ts}=\sum_{i=1}^n \left[\left\{-\frac{\exp[-2(1+\tau\eta_i)^{-1/\tau}]  (1+\tau\eta_i)^{2-\frac{2}{\tau}} }{\left(1- \exp[-(1+\tau\eta_i)^{-1/\tau}]\right)^2}
                 -\frac{ \exp[-(1+\tau\eta_i)^{-1/\tau}] (1+\tau\eta_i)^{2-\frac{2}{\tau}} }{1- \exp[-(1+\tau\eta_i)^{-1/\tau}]}+\right.\right. \\
                 &+& \left.\left. \frac{\exp[-(1+\tau\eta_i)^{-1/\tau}] (1+\tau)  (1+\tau\eta_i)^{2-\frac{1}{\tau}} }{1- \exp[-(1+\tau\eta_i)^{-1/\tau}]} \right\} (1-y_i)- (1+\tau) \left(1+\tau\eta_i\right)^{2-1/\tau} y_i\right]\B\ts_i\B_i.
\end{eqnarray*}
In practice, we use the more efficient and reliable trust region algorithm, which is based on (\ref{Fish}) \citep{Nocedal}.

The $\lambda_j$ are smoothing parameters controlling the trade-off between fit and smoothness, and in the above optimization problem they are fixed to some values. This is because joint estimation of $\bm\lambda\ts=(\lambda_1,\ldots,\lambda_p)$ and $\bm\delta$ via maximization of (\ref{pen.l}) would clearly lead to over-fitting since the highest value for $\ell_p(\bm\delta)$ would be obtained when $\bm\lambda=\textbf{0}$. In fact, $\bm\lambda$ should be selected so that the estimated smooth components are as close as possible to the true functions. Automatic multiple smoothing parameter selection can be achieved in several ways; see \cite{Ruppert:2003} and \cite{Wood:2006} for excellent overviews. Here, we elected to use a generalization of the approximate unbiased risk estimator \citep[UBRE,][]{CW79}. Specifically, $\hat{\bm\lambda}$ is the solution to the problem
\beq
\text{minimize \ \ } \frac{1}{n}\|{\sqrt{ {\bf W}} } (\mathbf{z}-\X\bm\delta)\|^2-1+\frac{2}{n}\text{tr}(\textbf{A}_{\bm\lambda}) \text{ \ \ w.r.t. \ \ }\bm\lambda,
\label{PIRLS1}
\eeq
where $\sqrt{{\bf W}}$ is a weight diagonal matrix square root, $\textbf{z}_i=\textbf{B}_i\ts \bm\delta^{[a]}+\textbf{W}_{[ii]}^{-1}\textbf{d}_i$, $\textbf{d}_i=\partial \ell(\bm\delta)_i/ \partial \eta_{i}$, $\textbf{W}_{[ii]}=-\partial^2 \ell(\bm\delta)_i/\partial \eta_{i}\partial \eta_{i}$, $\textbf{A}_{\bm\lambda}=\B(\B\ts{\bf W}\B+\textbf{S}_{\bm\lambda})^{-1}\B\ts{\bf W}$ is the hat matrix, $\B=(\B_1\ts,\ldots,\B_n\ts)\ts$ and $\text{tr}(\textbf{A}_{\bm\lambda})$ the estimated degrees of freedom ($edf$) of the penalized model. (Superscript $[a]$ has been suppressed from $\textbf{d}_i$, $\textbf{z}_i$ and $\textbf{W}_i$ to avoid clutter.) The working linear model quantities are constructed for a given estimate of $\bm\delta$, obtained from a trust region iteration. (\ref{PIRLS1}) will then produce an updated estimate for $\bm\lambda$ which will be used to obtain a new parameter vector estimate for $\bm\delta$. The two steps, one for $\bm\delta$ and the other for $\bm\lambda$, are iterated until convergence. Problem (\ref{PIRLS1}) is solved employing the approach by \cite{Wood2004} which can evaluate the approximate UBRE and its derivatives in a way that is both computationally efficient and stable.

As for the tail parameter in the GEV link, it is in principle possible to estimate jointly $\tau$ and $\bm\delta$ by maximizing (\ref{pen.l}). But this would complicate the parameter estimation process \cite[e.g.,][]{Smith85}. To keep things simple, we propose fitting as many BGEVA models as the number of a set of sensibly chosen values of $\tau$ and select the model that yields the best empirical predictive performance. In our case study, this proved to be a practical and effective means of handling this parameter.

\subsection{Confidence intervals and p-values\label{inference}}

Inferential theory for penalized estimators is complicated by the presence of smoothing penalties which undermines the usefulness of classic frequentist results \citep[e.g.,][]{Wood:2006}. As shown in \citet{MW12}, reliable point-wise confidence intervals for the terms of a model involving penalized regression splines can be constructed using
\beq
\bm\delta|{\bf y} \dot{\backsim} \mathcal{N}(\hat{\bm\delta},{\bf V}_{\bm\delta}),
\label{CBI}
\eeq
where ${\bf y}$ refers to the response vector, $\hat{\bm\delta}$ is an estimate of $\bm\delta$ and ${\bf V}_{\bm\delta} = (-\bm{\mathcal{J}}+{\bf S}_{\bm\lambda})^{-1}$. The structure of ${\bf V}_{\bm\delta}$ is such that it includes both a bias and variance component in a frequentist sense, a fact that makes such intervals have close to nominal coverage probabilities \citep{MW12}. Given (\ref{CBI}), confidence intervals for linear and non-linear functions of the model parameters can be easily obtained. For instance, for $\hat{f}_j(x_{ji})$ these can be obtained using
$$
\hat{f}_j(x_{ji}) \dot{\backsim} \mathcal{N}(f_j(x_{ji}),\textbf{B}_{ji}(x_{ji})\ts{\bf V}_{\bm\delta_j}\textbf{B}_{ji}(x_{ji})),
$$
where ${\bf V}_{\bm\delta_j}$ is the sub-matrix of ${\bf V}_{\bm\delta}$ corresponding to the regression spline parameters associated with $j^{th}$ smooth term, and $\textbf{B}_{ji}(x_{ji})\ts=[b_1(x_{ji}),\ldots,b_{K_j}(x_{ji})]$. For parametric model components, such as binary and categorical predictors, using (\ref{CBI}) is equivalent to using classic results because such terms are not penalized.

For smooth components, point-wise confidence intervals cannot be used for variable selection \citep[e.g.,][Chapter 6]{Ruppert:2003}. For this purpose, p-values can be employed. To construct them, we need the distribution of the $\textbf{f}_j$, where $\textbf{f}_j=\left[f_j(x_{j1}),f_j(x_{j2}),\ldots,f_j(x_{jn})\right]\ts$. As shown by \citet{W13}, in the context of extended generalized additive models, $\hat{\textbf{f}}_j\dot{\backsim} \mathcal{N}(\textbf{f}_j,{\bf V}_{\textbf{f}_j})$ where ${\bf V}_{\textbf{f}_j}=\textbf{B}_j(\textbf{x}_{j}){\bf V}_{\bm\delta_j}\textbf{B}_j(\textbf{x}_{j})\ts$ and $\textbf{B}_j(\textbf{x}_{j})$ denotes a full column rank matrix such that $\hat{\textbf{f}}_j=\textbf{B}_j(\textbf{x}_{j})\hat{\bm\gamma}_{j}$. It is then possible to obtain approximate p-values, for testing the hypothesis $\textbf{f}_j=\textbf{0}$, based on
$$
T_{r_j}=\hat{\textbf{f}}_j\ts {\bf V}_{\textbf{f}_j}^{r_j-}\hat{\textbf{f}}_j \dot{\backsim} \chi_{r_j}^2,
$$
where ${\bf V}_{\textbf{f}_j}^{r_j-}$ is the rank $r_j$ Moore-Penrose pseudo-inverse of ${\bf V}_{\textbf{f}_j}$, which can deal with rank deficiencies. Parameter $r_j$ is selected using the established notion of $edf$ employed in (\ref{PIRLS1}). Because $edf$ is not an integer, it can be rounded as follows \citep{W13}
$$
r_j=\left\{
\begin{gathered}
\text{floor}(edf_j) \ \ \ \ \ \ \ \ \ \ \text{if} \ \ edf_j < \text{floor}(edf_j) + 0.05 \hfill \\
\text{floor}(edf_j) + 1  \ \ \ \ \ \text{otherwise} \hfill \\
\end{gathered}
\right..
$$

\section{Case Study \label{EmpricalCase}}

The Italian SME sector is the largest in the EU, with 3,813 million of SMEs \citep{European:2012}. The vast majority of Italy's SMEs are micro-firms with less than 10 employees. In fact, Italy's share of micro-firms in all businesses, at $94.6\%$, exceeds the EU-average ($92.2\%$). Recovery from the financial crisis has been much weaker than that in other EU countries, and SMEs have been the hardest hit companies.

\subsection{Data}\label{data}

The empirical analysis is based on annual data for the period $2006-2011$ for $50,160$ Italian SMEs. The data are from AIDA-Bureau van Dijk, a large Italian financial and balance sheet information provider. The time horizon considered here is of some interest as it includes the financial crisis of 2008.

The definition of SME by the European Commission is adopted. That is, a business must have an annual turn-over of less than 50 million of Euros and the number of employees must not exceed 250 (\href{http://ec.europa.eu/enterprise/policies/sme/facts-figures-analysis/sme-definition/index.htm}{http://ec.europa.eu/enterprise/policies/sme/facts-figures-analysis/sme-definition/index.htm}). In this work, default is intended as the end of the SME's activity, that is the status in which the SME needs to liquidate its assets for the benefit of its creditors. In practice, we consider a default to have occurred when a specific SME enters a bankruptcy procedure as defined by the Italian law.

Following \cite{Altman:2006}, we applied choice-based or endogenous stratified sampling, where data are stratified based on the values of the response. Specifically, we drew randomly observations within each stratum defined by the two categories default and non-default, and considered all the defaulted firms. We then selected a random sample of non-defaulted SMEs (over the same years of defaults) to obtain a default percentage in our sample that was as close as possible to that for Italian SMEs, which is around $5 \%$ \citep{Cerved:2011}. Choice-based sampling makes observations dependent. Since the sample size is high, according to the super-population theory of \cite{Prentice:1986}, we assume that ours is a random sample \citep[e.g.,][]{Altman:2006}. We excluded firms for which information on accounting characteristics is not available.

Financial ratio analysis uses specific formulas to gain insight into a company and its operations. The main measures based on the balance sheet are the financial strength and activity ratios. The former provide information on how well a company is able to meet its obligations and how it is leveraged. This gives investors an idea of the financial stability of the company and its ability to finance itself. Activity ratios focus mainly on current accounts and show how well a company manages its operating cycle. These ratios provide insight into the operational efficiency of the company.

Since the exploratory variables may be highly dependent with each other, we carried out a multicollinearity analysis as done, for instance, in \cite{Altman:2006} and \cite{Ciampi:2008}. We discarded variables with a variance inflation factor (VIF) larger than 5 \citep[][pp. 257--258]{Greene:2000}. As a result, we eventually considered the 21 continuous covariates: number of employees, turnover per employee, long and medium term liabilities/total assets, current ratio, leverage, liquidity ratio, solvency ratio, banks/turnover, return on sales, return on equity, added value per employee, turnover/staff costs, return on shareholders' funds, return on capital employed, stock turnover ratio, profit per employee, average remuneration per year, pre-tax profit margin, cost of employees/added value, total liquid funds and debt/equity.

\subsection{Model fitting results}

\begin{figure}[htbp]
	\centering
		\includegraphics[width=0.5\textwidth,angle=270]{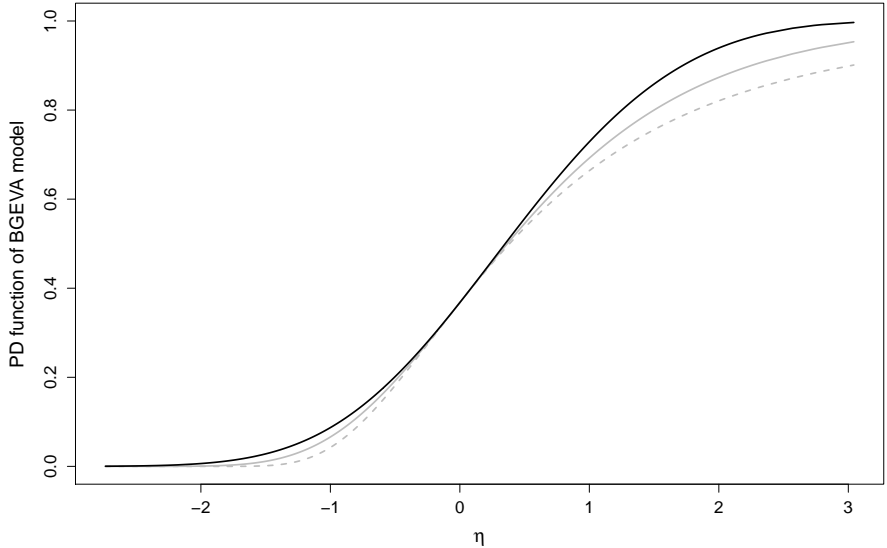}
\caption{$PD$ function of the BGEVA model for $\tau=-0.25$ (black solid line), $\tau=0.001$ (grey line) and $\tau=0.25$ (grey dashed line).
}
	\label{CDF}
\end{figure}

We employed the BGEVA model, logistic additive regression as well as log-log additive regression as it is a well known alternative to the logistic model. Computations were performed in the \texttt{R} environment \citep{CRAN} using the package \texttt{bgeva} \citep{bgeva} which can be used to fit the three models. The predictor specification was consistent with that shown in (\ref{SEMGEVA}). Smooth components were represented using penalized thin plate regression splines with basis dimensions equal to 20 and penalties based on second-order derivatives \citep{Wood:2006}. Using fewer or more spline bases did not lead to tangible changes in the results reported below and in the next section. Figure \ref{CDF} shows the $PD$ function of the BGEVA model (obtained from equation (\ref{SEMGEVA})) for $\tau$ equal to $-0.25$, $0.001$ (corresponding to the log-log response curve) and $0.25$. The skewness and approaching rate to 1 and to 0 depend on $\tau$. Compared with the log-log curve, for a negative value of $\tau$ the $PD$ of the BGEVA model approaches $0$ slowly but $1$ more sharply. Viceversa when $\tau$ is positive. Since in the context of our case study we need a curve which approaches 1 rapidly, negative values for $\tau$ are the obvious choice. For the BGEVA model, values of $\tau$ in the set $(-1.00, -0.75, -0.50, -0.25)$ were tried out. The best empirical predictive performance was produced for $\tau=-0.25$ (see next section). Using a finer grid of values and extending their range did not lead to improvements in terms of prediction.

As in \cite{Lin:2012}, \cite{Gumparthi:2010} and \cite{Fang:2011}, we carried out a stepwise procedure to identify a subset of predictors. After applying a backward selection procedure at the $5\%$ significance level, using the p-value definition discussed in Section \ref{inference}, only the 12 variables shown in Table \ref{TableREs} were kept. This selection result was consistent across all models.

\begin{figure}[htbp]
	\centering
		\includegraphics[width=1.1\textwidth,angle=270]{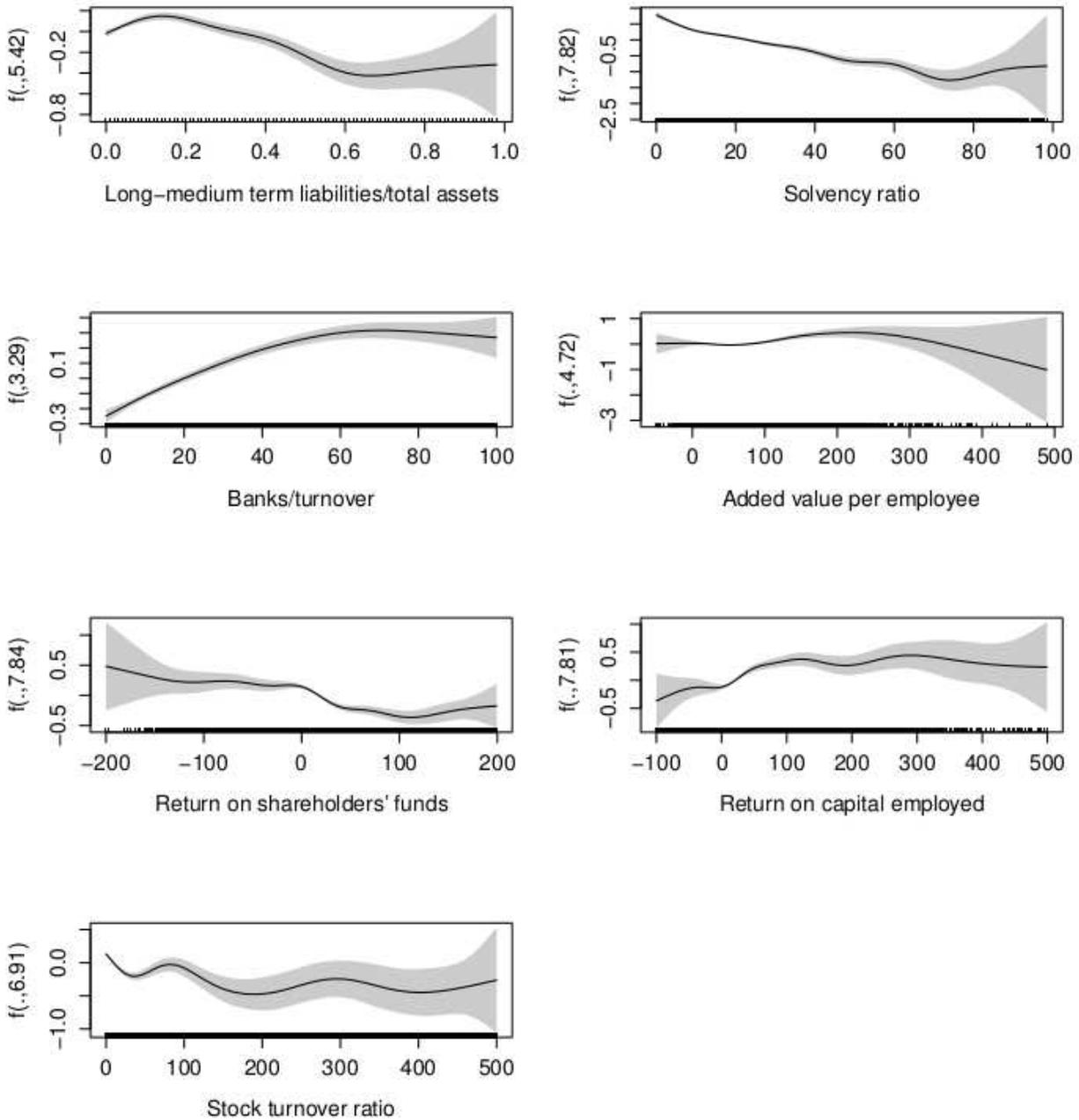}
\caption{Smooth component estimates of the 7 (out of 12) continuous variables that exhibit a non-linear pattern. These were obtained from applying the BGEVA model on the Italian SME data described in Section \ref{data}. Results are on the scale of the predictor. Shaded areas represent $95\%$ confidence intervals and the ‘rug plot’, at the bottom of each graph, is used to show the covariate values. The numbers in brackets in the y-axis captions are the estimated degrees of freedom ($edf$) of the smooth curves.}
	\label{plots}
\end{figure}

Some of the covariate effects are reported in the parametric part of the BGEVA model since their smooth function estimates were linear (i.e. their $edf$s were all equal to 1). For the other explanatory variables, the estimated smooths exhibit $edf$s considerably greater than 1 and are displayed in Figure \ref{plots}.

\begin{table}
\begin{center}
\begin{tabular}{|l|c|c|c|c|}
\hline
\emph{Variable names of parametric model part} & \emph{Estimate} &  \emph{Std. Error} & \emph{z value} & \emph{p-value}\\
  \hline
Intercept                  & -1.675 & 2.926e-02 & -57.247 &  $<$ 2e-16\\
Leverage                   & 0.001  & 4.642e-04 & 2.700   & 0.006943 \\
Turnover/staff costs       & 0.004  & 1.061e-03 & 3.428   & 0.000607 \\
Profit per employee        & -0.009 & 1.126e-03 & -7.697  & 1.39e-14\\
Cost of employees/added value & 0.004  & 8.835e-04 & 4.902   & 9.46e-07 \\
Total liquid funds         & 0.001   & 1.231e-06 & 117.892 & $<$ 2e-16 \\
 \hline
\emph{Variable names of smooth terms} & \emph{Edf} &  \emph{Est.rank} & \emph{Chi.sq} & \emph{p-value}\\
  \hline
Long-medium term liabilities/total assets & 5.42  & 6 & 124.70  & $<$ 2e-16 \\
Solvency ratio                            & 7.82  & 8 & 123.78  & $<$ 2e-16 \\
Banks/turnover 														& 3.29  & 4 & 37.33  & 1.54e-07\\
Added value per employee 									& 4.72  & 5 & 21.64  & 0.000613\\
Return on shareholders' funds 						& 7.84  & 8 & 106.09  & $<$ 2e-16 \\
Return on capital employed 								& 7.81  & 8 & 71.22  & 2.81e-12\\
Stock turnover ratio  										& 6.91  & 7 & 92.48  & $<$ 2e-16 \\
 \hline
\end{tabular}
\caption{Parametric and smooth component summaries obtained from applying the BGEVA model to a sample of Italian SMEs (2,384 defaulters and 45,296 non-defaulters) for the period $2006-2011$. Full details on the calculation of the test statistic for the smooth terms are given in Section \ref{inference}.}
\label{TableREs}
\end{center}
\end{table}

Analogously to logistic model (\ref{logit}), if the estimated coefficient of a covariate is positive, an increase in its value will result in a increase in the estimated $PD$, holding constant all the other variables. Viceversa when the coefficient is negative. The leverage result in Table \ref{TableREs} is consistent with financial theory. Financially distressed SMEs would be expected to have lower value for this variable than healthier firms would \citep{Altman:2010}. Turnover/staff costs shows a positive relationship with $PD$ which is in contrast with \cite{Ciampi:2013} who considered turnover per employee. Profit per employee is in agreement with the expectations negatively related to the likelihood of being a defaulted SME. 

 Cost of employees/added value shows a positive relationship with $PD$ which is also coherent with the literature: \cite{Ciampi:2008} for total personnel costs/added value and \cite{Fantazzini:2009} for personnel expenses/sales. To explain the positive relationship between total liquid funds and $PD$, we refer to an interesting European Commission paper by \cite{Ferrando:2007}. Here, liquid funds are higher if there is a higher probability of a shortage of funds in the future that could be due to financial distress in SMEs.

As for the plots in Figure \ref{plots}, in line with the interpretation for the parametric components, if the estimated smooth function of a covariate is decreasing then the estimated $PD$ decreases, and viceversa. An interesting result is obtained for bank loans/turnover. For low values of this covariate the relationship to $PD$ is positive: for low debt loans, bank loans decrease SME's solvency. When the debt load is too high, an increase in bank loans to an SME results in a slight (perhaps not significant) decrease in the $PD$. Instead, \cite{Ciampi:2008} found a positive relationship between bank loans/turnover and $PD$. Alternatively, \cite{Altman:2006} considered short term debt/equity book value to model $PD$ and found an inverse relationship. On the contrary, \cite{Fantazzini:2009} showed that short term debt has a positive influence on $PD$, which is consistent with expectations. These results are not all in agreement. This is probably because these authors modeled covariate effects assuming linearity whereas we employed smooth functions which allowed us to estimate flexibly such effects. 

Another interesting result is obtained for long-medium term liabilities/total assets: for low values this variable shows a direct relationship to $PD$ which then becomes negative for high covariate values. When the ratio is higher than 0.7, the estimated  curve is slightly increasing although the confidence intervals are wide and hence it is not possible to be conclusive. The positive impact of long-medium term liabilities/total assets to PD for values lower than 1.7 and higher than 0.7 may justify the results obtained by some authors who found a positive impact assuming linearity \citep[e.g.,][]{Arslan:2009}.

For values of added value per employee lower than about 200 thousands of Euros, the relationship to $PD$ is slightly positive (almost constant), similar to the results obtained by \cite{Ciampi:2008}. For high values of this covariate, Figure \ref{plots} shows a negative relationship and coherent with expectations, although the confidence intervals are wide and contain the case of an inverse relationship as well. A few analyses have included return on capital employed as an explanatory variable in scoring models for SMEs \citep[e.g.,][]{Toyli:2008}. This covariate shows a positive relationship with the likelihood of being a defaulted SME, coherently with expectations, except for high covariate values where the confidence intervals are so wide that many interpretations are plausible. As for the other variables, a clear interpretation can not be provided and further research is needed. The estimates of the parametric and smooth components obtained by applying the logistic and log-log additive models (not reported here) did not differ significantly from each other and are in agreement with those presented in this section.

\subsection{Empirical predictive performance}

In this section, we compare the predictive performance of the BGEVA, logistic and log-log additive models. Predictive accuracy can be assessed using the Mean Squared Error (MSE) and Mean Absolute Error (MAE) based on observed default and predicted $PD$. Scoring models with lower MSE and MAE should forecast defaults and non-defaults more accurately. However, it is the identification of defaulters that is a pivotal aim for banks' internal scoring models. In fact, it is much more costly to classify an SME as a non-defaulter when it is a defaulter than to classify it as a defaulter when it is not. If a defaulted firm is classified as a non-defaulter by the scoring model, then the bank will give a loan. If the borrower becomes defaulted, then the bank may lose the whole or a part of its credit exposure. On the contrary, when a non-defaulter is classified as a defaulter, the bank only loses interest on loans. We, therefore, computed the MAE and MSE only for defaulted SMEs and denote them by MAE$^+$ and MSE$^+$. We also calculated the area under the curve (AUC) index \citep{Hand:2001} and H-measure \citep{Hand:2010}, using the \texttt{R} package \texttt{hmeasure} with a severity ratio of 0.01 for the H-measure. Scoring models with higher AUC and H-measure can forecast defaults and non-defaults more accurately. Note that the H-measure overcomes the drawbacks of AUC when the class sizes and classification error costs are extremely unbalanced \citep{Hand:2009,Hand:2010}, which is especially true for credit scoring applications.

To avoid sample dependency, models were validated on observations that were not included in the sample used to estimate the model. Specifically, we used out-of-sample and out-of-time tests. In the former, the models were estimated using all observations in $2006-2011$ but a randomly drawn control sample of size corresponding to the $10\%$ of all observations. In the latter, we analyzed the models' performance for different sample sizes and default horizons. This was achieved in two ways. In the first case, models were fitted using data from the intervals $2006-2010$, $2006-2009$ and $2006-2008$ and their predictive accuracy evaluated on SMEs belonging to $2011$, $2010$ and $2009$, respectively. In the second case, we estimated the models using data for the period $2006-2008$ and tested them on SMEs belonging to $2009-2010$ and $2009-2011$. 
 The estimation period of $2006-2008$ was chosen because it was characterized by an economic growth in Italy which was then followed by a economic crisis after $2008$.

\begin{table}
\begin{center}

\begin{tabular}{|l|c|c|c|c|}
\hline
\emph{Type of control sample} & \hspace{1mm}\emph{measure} & \emph{BGEVA} & \emph{log-log} & \emph{logistic}\\

\hline

 \multirow{4}{*}{\vspace{8mm}\emph{Out-of-sample}} & MAE$^+$ & 0.754 & 0.791  & 0.889\\
                                                   & MSE$^+$ & 0.626 & 0.660 & 0.795\\
                                                   &  H      & 0.184 & 0.050  & 0.018\\
                                                   &  AUC    & 0.682 & 0.680 & 0.811\\
                                                   \hline
  \multirow{4}{*}{\vspace{8mm}\emph{Out-of-time $2009$}} & MAE$^+$ &0.837  &0.849  &0.883 \\
                                                         &  MSE$^+$ &0.741  &0.754  &0.786 \\
                                                         &  H      &0.136  &0.052  & 0.020\\
                                                         &  AUC    &0.697  &0.692  & 0.809 \\

   \hline
  \multirow{4}{*}{\vspace{8mm}\emph{Out-of-time $2010$}} & MAE$^+$ & 0.770 &0.798  &0.881 \\
                                                              &  MSE$^+$ &0.620&0.657 & 0.783\\
                                                              &  H & 0.081& 0.078& 0.020\\
                                                              &  AUC & 0.722& 0.722& 0.811\\
    \hline
  \multirow{4}{*}{\vspace{8mm}\emph{Out-of-time $2011$}} & MAE$^+$ & 0.843 &0.862  &0.894 \\
                                                         &  MSE$^+$ & 0.763&0.781 & 0.804\\
                                                         &  H & 0.050 & 0.027 &0.030 \\
                                                         &  AUC &0.619 & 0.599 &0.808 \\
   \hline

 \end{tabular}
\caption{Forecasting accuracy measures for out-of-sample and out-of-time exercises obtained from applying the BGEVA, log-log and logistic additive models. Default horizon is of one year.}
\label{Tabl3}
\end{center}
\end{table}

Table \ref{Tabl3} reports the values of MAE$^+$, MSE$^+$, AUC index and H-measure for the out-of-sample and out-of-time exercises when the default horizon is of one year. The MAE$^+$ and MSE$^+$ for the BGEVA model are lower than those for the log-log and logistic additive models. The AUC of the BGEVA and log-log models are lower than that of logistic regression in all control samples. However, as pointed out earlier, AUC is not reliable when the class sizes and classification error costs are unbalanced \citep{Hand:2009,Hand:2010}, case in which the H-measure is more appropriate \citep{Hand:2010}. The H-measure for BGEVA is higher than that of its competitors in all control samples. Increasing the default horizon does not lead to different conclusions (see Table \ref{Tabl4}). The forecasting periods considered here are important since the effects of the financial crisis on Italian SMEs have been strong in 2009, where the number of defaulted Italian SMEs in the sample was very high (701), and decreased in 2010 and 2011 (429 and 124, respectively).

\begin{table}
\label{tab:1}
\begin{center}
\begin{tabular}{|l|c|c|c|c|}
\hline
\emph{Type of control sample} & \hspace{1mm}\emph{measure} & \emph{BGEVA} & \emph{log-log} & \emph{logistic}\\

\hline

   \hline
  \multirow{4}{*}{\vspace{8mm}\emph{Two years: $2009-2010$}}      & MAE$^+$ & 0.848 &0.862  &0.882 \\
                                                                   &  MSE$^+$ &0.746 &0.760 & 0.785\\
                                                                   &  H & 0.188& 0.066& 0.019\\
                                                                   &  AUC & 0.712& 0.711& 0.811\\
   \hline
  \multirow{4}{*}{\vspace{8mm}\emph{Three years: $2009-2011$}} & MAE$^+$ & 0.832 &0.845  &0.881 \\
                                                                   &  MSE$^+$ &0.724&0.740 & 0.783\\
                                                                   &  H & 0.190& 0.067& 0.019\\
                                                                 &  AUC & 0.713& 0.714& 0.811\\
                                                                  \hline
    \end{tabular}
\caption{Forecasting accuracy measures for out-of-sample and out-of-time exercises obtained from applying the BGEVA, log-log and logistic additive models. Default horizons are of two and three years.}
\label{Tabl4}
\end{center}
\end{table}

Our analysis suggests that the accounting-based BGEVA model has a superior predictive performance in identifying defaulted SMEs than that of its traditional alternatives. Since identification of defaulters is crucial for banks, the proposed tool could be employed as their internal scoring model to identify defaulted SMEs.

\section{Concluding remarks}

We introduced a scoring model to forecast defaulted SMEs. Since lending to SMEs is risky and because they play a crucial role in the economy of many countries, we would expect the BGEVA model to be generally useful for analyzing SMEs.

The proposed approach is based on a penalized likelihood estimation framework where the link function is allowed to be asymmetric (through the use of the quantile function of a GEV random variable) and the functional dependence of the response on continuous covariates is flexibly determined from the data (through the use of penalized regression splines). The developments discussed in this article are implemented in the \texttt{R} package \texttt{bgeva} \citep{bgeva} which can be particularly attractive to researchers and practitioners who wish to fit BGEVA models.

Our proposal and its competitors were applied to data on 50,160 Italian SMEs for the period $2006-2011$. The empirical results confirmed that the first main advantage of the BGEVA model lies in its superior performance in forecasting defaulted SMEs for different default horizons. The second is the relaxation of the linearity assumption which was not clearly supported by the data.

Banks and financial institutions could improve their internal assessments and efficiency by using BGEVA models in that they can better identify defaulted SMEs and can shed light on the nature of the relationships between response and SME characteristics. It would be interesting to explore the impact that non-random sample selection \citep{Banasik07} has on bankruptcy prediction of SMEs and future research will look at the possibility of extending the BGEVA model to account for selection bias.

\bibliographystyle{apalike2}
\bibliography{Ref}

\begin{thebibliography}{}

\bibitem[Agresti, 2002]{Agresti:2002}
Agresti, A. (2002).
\newblock {\em Categorical Data Analysis}.
\newblock Wiley, New York.

\bibitem[Altman, 1968]{Altman:1968}
Altman, E. (1968).
\newblock Financial ratios, discriminant analysis and the prediction of
  corporate bankruptcy.
\newblock {\em Journal of Finance}, 23(4), 589--609.

\bibitem[Altman \& Sabato, 2006]{Altman:2006}
Altman, E. \& Sabato, G. (2006).
\newblock Modeling credit risk for smes: Evidence from the us market.
\newblock {\em ABACUS}, 19(6), 716--723.

\bibitem[Altman et~al., 2010]{Altman:2010}
Altman, E., Sabato, G., \& Wilson, N. (2010).
\newblock The value of non-financial information in small and medium-sized
  enterprise risk management.
\newblock {\em The Journal of Credit Risk}, 6, 1--33.

\bibitem[Arslan, 2009]{Arslan:2009}
Arslan, O. M.~B. (2009).
\newblock Credit risks and internationalization of smes.
\newblock {\em Journal of Business Economics and Management}, 10(4), 361--368.

\bibitem[Banasik \& Crook, 2007]{Banasik07}
Banasik, J. \& Crook, J. (2007).
\newblock Reject inference, augmentation, and sample selection.
\newblock {\em European Journal of Operational Research}, 183, 1582--1594.

\bibitem[{BCBS}, 2005]{Basel:2005}
{BCBS} (2005).
\newblock International convergence of capital measurement and capital
  standards: A revised framework.
\newblock {\em Bank for International Settlements}, 25.

\bibitem[Becchetti \& Sierra, 2002]{Becchetti:2002}
Becchetti, L. \& Sierra, J. (2002).
\newblock Bankruptcy risk and productive efficiency in manufacturing firms.
\newblock {\em Journal of Banking and Finance}, 27, 2099--2120.

\bibitem[Berg, 2007]{Berg:2007}
Berg, D. (2007).
\newblock Bankruptcy prediction by generalized additive models.
\newblock {\em Applied Stochastic Models in Business and Industry}, 23,
  129--143.

\bibitem[Berger \& Udell, 2002]{Berger:2002}
Berger, A.~N. \& Udell, G.~F. (2002).
\newblock Small business credit availability and relationship lending: The
  importance of bank organisational structure.
\newblock {\em Finance and Economics Discussion Series paper}, 23, 2004--2012.

\bibitem[Berger \& Udell, 2006]{Berger:2006}
Berger, A.~N. \& Udell, G.~F. (2006).
\newblock A more complete conceptual framework for sme finance.
\newblock {\em Journal of Banking and Finance}, 30, 2945--2966.

\bibitem[Calabrese \& Osmetti, 2013]{Calabrese:2013}
Calabrese, R. \& Osmetti, S.~A. (2013).
\newblock Modelling sme loan defaults as rare events: the generalized extreme
  value regression model.
\newblock {\em Journal of Applied Statistics}, 40(6), 1172--1188.

\bibitem[Cerved-Group, 2011]{Cerved:2011}
Cerved-Group (2011).
\newblock Caratteristiche delle imprese, governance e probabilit\`{a} di
  insolvenza.
\newblock In {\em Report, Milan, IT}.

\bibitem[Chuang \& Lin, 2009]{Chuang:2009}
Chuang, C.~L. \& Lin, R.~H. (2009).
\newblock Constructing a reassigning credit scoring model.
\newblock {\em Expert Systems with Applications}, 36, 1685--1694.

\bibitem[Ciampi \& Gordini, 2008]{Ciampi:2008}
Ciampi, F. \& Gordini, N. (2008).
\newblock Using economic-financial ratios for small enterprise default
  prediction modeling: An empirical analysis.
\newblock In {\em Proceedings of the Oxford Business $\&$ Economics Conference,
  Oxford, UK}  (pp.\ 1--21).

\bibitem[Ciampi \& Gordini, 2013]{Ciampi:2013}
Ciampi, F. \& Gordini, N. (2013).
\newblock Small enterprise default prediction modelling through artificial
  neural networks: an empirical analysis of italian small enterprises.
\newblock {\em Journal of Small Business Management}, 51(1), 23--45.

\bibitem[Craven \& Wahba, 1979]{CW79}
Craven, P. \& Wahba, G. (1979).
\newblock Smoothing noisy data with spline functions.
\newblock {\em Numerische Mathematik}, 31, 377--403.

\bibitem[Dietsch \& Petey, 2004]{Dietsch:2004}
Dietsch, M. \& Petey, J. (2004).
\newblock Should sme exposure be treated as retail or as cor- porate exposures?
  a comparative analysis of default probabilities and asset correlation in
  french and german smes.
\newblock {\em Journal of Banking and Finance}, 28, 773--788.

\bibitem[EC, 2012]{European:2012}
EC (2012).
\newblock Sba fact sheet 2012 for italy.
\newblock {\em Enterprise and Industry working paper}.

\bibitem[Eilers \& Marx, 1996]{EM96}
Eilers, P. H.~C. \& Marx, B.~D. (1996).
\newblock Flexible smoothing with {B}-splines and penalties.
\newblock {\em Statistical Science}, 11(2), 89--121.

\bibitem[Falk et~al., 2010]{Falk:2010}
Falk, M., Haler, J., \& Reiss, R. (2010).
\newblock {\em Laws of Small Numbers: Extremes and Rare Events}.
\newblock Springer.

\bibitem[Fang \& Huang, 2011]{Fang:2011}
Fang, K. \& Huang, H. (2011).
\newblock Variable selection for credit risk model using data mining technique.
\newblock {\em JOURNAL OF COMPUTERS}, 6(9), 1868--1874.

\bibitem[Fantazzini \& Figini, 2009]{Fantazzini:2009}
Fantazzini, D. \& Figini, S. (2009).
\newblock Random survival forests models for sme credit risk measurement.
\newblock {\em Methodology and Computing in Applied Probability}, 11, 29--45.

\bibitem[Ferrando et~al., 2007]{Ferrando:2007}
Ferrando, A., Kohler-Ulbrich, P., \& Pal, R. (2007).
\newblock Is the growth of euro area small and medium-sized enterprises
  constrained by financing barriers?
\newblock {\em Industrial Policy and Economic Reforms, Enterprise and Industry
  Directorate, General European Commission}, (6).

\bibitem[Gestel et~al., 2005]{Gestel:2005}
Gestel, T.~V., Baesens, B., Dijcke, P.~V., Suykens, J. A.~K., Garcia, J., \&
  Alderweireld, T. (2005).
\newblock Linear and non-linear credit scoring by combining logistic regression
  and support vector machines.
\newblock {\em Journal of Credit Risk}, 1(4), 31--60.

\bibitem[Greene, 2012]{Greene:2000}
Greene, W.~H. (2012).
\newblock {\em Econometric Analysis}.
\newblock Prentice Hall, New York.

\bibitem[Gumparthi \& Manickavasagam, 2010]{Gumparthi:2010}
Gumparthi, S. \& Manickavasagam, V. (2010).
\newblock Risk classification based on discriminant analysis for smes’.
\newblock {\em International Journal of Trade, Economics and Finance}, 1(3),
  242--246.

\bibitem[Gupton et~al., 1997]{Gupton:1997}
Gupton, G.~M., Finger, C.~C., \& Bhatia, M. (1997).
\newblock Creditmetrics.
\newblock In {\em Technical document}: J. P. Morgan.

\bibitem[Hand, 2009]{Hand:2009}
Hand, D.~J. (2009).
\newblock Measuring classifier performance: a coherent alternative to the area
  under the roc curve.
\newblock {\em Machine Learning}, 77, 103--123.

\bibitem[Hand, 2010]{Hand:2010}
Hand, D.~J. (2010).
\newblock Evaluating diagnostic tests: the area under the roc curve and the
  balance of errors.
\newblock {\em Statistics in Medicine}, 29, 1502--1510.

\bibitem[Hand \& Henley, 1997]{Hand:1997}
Hand, D.~J. \& Henley, W.~E. (1997).
\newblock Some developments in statistical credit scoring.
\newblock In N. Nakhaeizadeh \& C. Taylor (Eds.), {\em Machine learning and
  statistics: the interface}  (pp.\ 221--237).: Wiley, New York.

\bibitem[Hand et~al., 2001]{Hand:2001}
Hand, D.~J., Mannila, H., \& Smyth, P. (2001).
\newblock {\em Principles of Data Mining}.
\newblock MIT Press.

\bibitem[Huang et~al., 2006]{Huang:2006}
Huang, J.~J., Tzeng, J.~H., \& Ong, C.~S. (2006).
\newblock Two-stage genetic programming (2sgp) for the credit scoring model.
\newblock {\em Applied Mathematics and Computation}, 174, 1039--1053.

\bibitem[Jarrow \& Turnbull, 1995]{Jarrow:1995}
Jarrow, R.~A. \& Turnbull, S.~M. (1995).
\newblock Pricing derivatives on financial securities subject to credit risk.
\newblock {\em Journal of Finance}, 50, 53--85.

\bibitem[Kiefer, 2010]{Kiefer:2010}
Kiefer, N.~M. (2010).
\newblock Journal of business and economic statistics.
\newblock {\em Journal of Business Finance \& Accounting}, 28(2), 320--328.

\bibitem[King \& Zeng, 2001]{King:2001}
King, G. \& Zeng, L. (2001).
\newblock Logistic regression in rare events data.
\newblock {\em Political Analysis}, 9, 321--354.

\bibitem[Kotz \& Nadarajah, 2000]{Kotz:2000}
Kotz, S. \& Nadarajah, S. (2000).
\newblock {\em Extreme Value Distributions. Theory and Applications}.
\newblock Imperial College Press, London.

\bibitem[Lee \& Chen, 2005]{Lee:2005}
Lee, T.~S. \& Chen, I.~F. (2005).
\newblock A two-stage hybrid credit scoring model using artificial neural
  networks and multivariate adaptive regression splines.
\newblock {\em Expert Systems with Applications}, 28, 743--752.

\bibitem[Lin et~al., 2012]{Lin:2012}
Lin, S.~M., Ansell, J., \& Andreeva, G. (2012).
\newblock Predicting default of a small business using different definitions of
  financial distress.
\newblock {\em Journal of the Operational Research Society}, 63, 539--548.

\bibitem[Marra et~al., 2013]{bgeva}
Marra, G., Calabrese, R., \& Osmetti, S.~A. (2013).
\newblock {\em bgeva: Binary Generalized Extreme Value Additive Models}.
\newblock R package version 0.2.

\bibitem[Marra \& Radice, 2010]{MR10}
Marra, G. \& Radice, R. (2010).
\newblock Penalised regression splines: Theory and application to medical
  research.
\newblock {\em Statistical Methods in Medical Research}, 19, 107--125.

\bibitem[Marra \& Wood, 2012]{MW12}
Marra, G. \& Wood, S. (2012).
\newblock Coverage properties of confidence intervals for generalized additive
  model components.
\newblock {\em Scandinavian Journal of Statistics}, 39, 53--74.

\bibitem[McCullagh \& Nelder, 1989]{McCullagh:1989}
McCullagh, P. \& Nelder, J. (1989).
\newblock {\em Generalized Linear Models}.
\newblock Chapman Hall, New York.

\bibitem[Merton, 1974]{Merton:1974}
Merton, R. (1974).
\newblock On the pricing of corporate debt: The risk structure of interest
  rates.
\newblock {\em Journal of Finance}, 29, 449--470.

\bibitem[Nocedal \& Wright, 2006]{Nocedal}
Nocedal, J. \& Wright, S.~J. (2006).
\newblock {\em Numerical Optimization}.
\newblock Springer-Verlag, New York.

\bibitem[Ong et~al., 2005]{Ong:2005}
Ong, C.~S., Huanga, J.~J., \& Tzeng, G.~H. (2005).
\newblock Building credit scoring models using genetic programming. expert
  systems with applications.
\newblock {\em Expert Systems with Applications}, 29, 41--47.

\bibitem[Pederzini, 2012]{Pederzini:2012}
Pederzini, E. (2012).
\newblock European policies to promote the access to finance of smes.
\newblock In C.~G. B.~Dallago (Ed.), {\em The Consequences of the International
  Crisis on European SMEs - Vulnerability and Resilience}  (pp.\ 89--106).:
  Routledge.

\bibitem[Prentice, 1986]{Prentice:1986}
Prentice, R.~L. (1986).
\newblock A case-cohort design for epidemiologic cohort studies and disease
  prevention trials.
\newblock {\em Biometrika}, 66, 403--411.

\bibitem[Products, 1997]{Credit:1997}
Products, C. S.~F. (1997).
\newblock Creditrisk+: A credit risk management framework.
\newblock In {\em Credit Suisse First Boston}.

\bibitem[{R Development Core Team}, 2013]{CRAN}
{R Development Core Team} (2013).
\newblock {\em R: A Language and Environment for Statistical Computing}.
\newblock R Foundation for Statistical Computing, Vienna, Austria.
\newblock {ISBN} 3-900051-07-0.

\bibitem[Ruppert et~al., 2003]{Ruppert:2003}
Ruppert, D., Wand, M.~P., \& Carroll, R.~J. (2003).
\newblock {\em Semiparametric Regression}.
\newblock Cambridge University Press, London.

\bibitem[Saurina \& Trucharte, 2004]{Saurina:2004}
Saurina, J. \& Trucharte, C. (2004).
\newblock The impact of basel ii on lending to small- and medium-sized firms: A
  regulatory policy assessment based on span- ish credit register data.
\newblock {\em Journal of Finance Service Research}, 26(3), 121--144.

\bibitem[Smith, 1985]{Smith85}
Smith, R.~L. (1985).
\newblock Maximum likelihood estimation in a class of non-regular cases.
\newblock {\em Biometrika}, 72, 67�--90.

\bibitem[Thomas et~al., 2002]{Thomas:2004}
Thomas, L., Edelman, D., \& Crook, J.~C. (2002).
\newblock {\em Credit Scoring and Its Applications}.
\newblock Society for Industrial and Applied Mathematics, Philadelphia.

\bibitem[Toyli et~al., 2008]{Toyli:2008}
Toyli, J., Hakkinen, L., Ojala, L., \& Naula, T. (2008).
\newblock Logistics and financial performance. an analysis of 424 finnish small
  and medium-sized enterprises.
\newblock {\em International Journal of Physical Distribution $\&$ Logistics
  Management}, 38(1), 57--80.

\bibitem[Wang \& Dey, 2010]{Wang2010}
Wang, X. \& Dey, D.~K. (2010).
\newblock Generalized extreme value regression for binary response data: An
  application to b2b electronic payments system adoption.
\newblock {\em The Annals of Applied Statistics}, 4(4), 2000--2023.

\bibitem[Wilson, 1998]{Wilson:1998}
Wilson, T.~C. (1998).
\newblock The impact of basel ii on lending to small-and medium-sized firms: A
  regulatory policy assessment based on spanish credit register data.
\newblock {\em Economic Policy Review}, 4, 71--82.

\bibitem[Wood, 2004]{Wood2004}
Wood, S.~N. (2004).
\newblock Stable and efficient multiple smoothing parameter estimation for
  generalized additive models.
\newblock {\em Journal of the American Statistical Association}, 99, 673--686.

\bibitem[Wood, 2006]{Wood:2006}
Wood, S.~N. (2006).
\newblock {\em Generalized Additive Models: An Introduction with R}.
\newblock Chapman $\&$ Hall, Boca Raton.

\bibitem[Wood, 2013]{W13}
Wood, S.~N. (2013).
\newblock On p-values for smooth components of an extended generalized additive
  model.
\newblock {\em Biometrika}, 100, 221--228.

\bibitem[Zavgren, 1998]{Zavgren:1998}
Zavgren, C. (1998).
\newblock The prediction of corporate failure: the state of the art.
\newblock {\em Journal of Accounting Literature}, 2, 1--37.

\end{thebibliography}

\end{document}